\documentclass[conference]{IEEEtran}
\IEEEoverridecommandlockouts

\usepackage{cite}
\usepackage{amsmath,amssymb,amsfonts}
\usepackage{algorithmic}
\usepackage{graphicx}
\usepackage{textcomp}
\usepackage{xcolor}

\usepackage{amsmath, calc}
\usepackage{graphicx}
\usepackage{tcolorbox}
\usepackage{multirow}
\usepackage{url}
\usepackage{hyperref}
\usepackage{pgfplots}
\pgfplotsset{compat=1.18}
\usepackage{textcomp}
\usepackage{placeins}
\usepackage{fontawesome}
\usepackage{comment}
\usepackage{tabularx}
\usepackage{enumitem}
\usepackage{booktabs}
\usepackage{comment}

\usepackage{hyperref}

\usepackage{framed}

\usepackage{tcolorbox}
\usepackage{fancybox} 

\usepackage{tabularx}

\usepackage{makecell}
\usepackage{multirow}
\usepackage{csquotes}
\usepackage{graphicx}
\usepackage{orcidlink}
\usepackage{multicol}
\usepackage{colortbl}
\usepackage{fancybox,framed}
\usepackage{soul}

\def\BibTeX{{\rm B\kern-.05em{\sc i\kern-.025em b}\kern-.08em
    T\kern-.1667em\lower.7ex\hbox{E}\kern-.125emX}}
\begin{document}

\makeatletter
\newcommand{\linebreakand}{%
  \end{@IEEEauthorhalign}
  \hfill\mbox{}\par
  \mbox{}\hfill\begin{@IEEEauthorhalign}
}
\makeatother

\title{A German Gold-Standard Dataset for Sentiment Analysis in Software Engineering}


\author{
    \IEEEauthorblockN{
        Martin Obaidi\orcidlink{0000-0001-9217-3934}, 
        Marc Herrmann\orcidlink{0000-0002-3951-3300}, 
        Elisa Schmid\orcidlink{0009-0006-2498-9986},\\
        Raymond Ochsner,
        Kurt Schneider\orcidlink{0000-0002-7456-8323}
    }
    \IEEEauthorblockA{
        \textit{Leibniz Universität Hannover, Software Engineering Group} \\
        Hannover, Germany \\
        \{martin.obaidi, marc.herrmann\}@inf.uni-hannover.de, \\
        \{elisa.schmid, kurt.schneider\}@inf.uni-hannover.de, \\ raymond.ochsner@stud.uni-hannover.de
    }
    \and
    \IEEEauthorblockN{
        Jil Klünder\orcidlink{0000-0001-7674-2930}
    }
    \IEEEauthorblockA{
        \textit{University of Applied Sciences} \\
        \textit{FHDW Hannover} \\
        Hannover, Germany \\
        jil.kluender@fhdw.de
    }
}


\maketitle

\begin{abstract}
Sentiment analysis is an essential technique for investigating the emotional climate within developer teams, contributing to both team productivity and project success. Existing sentiment analysis tools in software engineering primarily rely on English or non-German gold-standard datasets. To address this gap, our work introduces a German dataset of 5,949 unique developer statements, extracted from the German developer forum Android-Hilfe.de. Each statement was annotated with one of six basic emotions, based on the emotion model by Shaver et al.~\cite{shaver1987emotion}, by four German-speaking computer science students. Evaluation of the annotation process showed high interrater agreement and reliability. These results indicate that the dataset is sufficiently valid and robust to support sentiment analysis in the German-speaking software engineering community. Evaluation with existing German sentiment analysis tools confirms the lack of domain-specific solutions for software engineering. We also discuss approaches to optimize annotation and present further use cases for the dataset.

\end{abstract}

\begin{IEEEkeywords}
sentiment analysis, requirements engineering, developer communication. machine learning
\end{IEEEkeywords}

\section{Introduction}
\label{sec:intro}

Communication among developers plays a central role in software engineering and has a major impact on both productivity and project success~\cite{chepenik2007influence, herrmann2025different-perceptions, graziotin2014happy}. In online developer communities and chatrooms, countless discussion threads are created every day, where developers ask questions, provide answers, and exchange views on technical challenges~\cite{pan2021automatingdeveloperchat, ehsan2021developerdiscussion}.

Moods have a significant impact on our cognitive processes and, consequently, on our behavior~\cite{herrmann2025montecarlo,graziotin2014happy}. This influence is particularly pronounced in software engineering (SE) teams, where expressed moods shape interactions and can directly affect team members~\cite{graziotin2014happy, herrmann-kluender-2021, herrmann2021automatic}. Graziotin et al.~\cite{graziotin2014happy} found that positive moods within developer teams lead to increased productivity.

In the context of Crowd-Based Requirements Engineering (CrowdRE), analyzing developer communication is essential for understanding both technical and social dynamics within large, distributed communities. Sentiment analysis of developer discussions enables the identification of software components or topics associated with negative emotions, highlighting areas that may require special attention during requirements elicitation~\cite{tulili2024characterisingdevelopersentimentsoftware, tulili2025developersentiment}. Tulili et al.~\cite{tulili2025developersentiment} have shown that developers working on more complex components tend to express negative sentiments more frequently.

Furthermore, tracking sentiment trends supports the early detection of potential conflicts~\cite{guilherme2016trustsentiment} and provides valuable insights into team culture and productivity~\cite{miller2021productivity}, as well as explainability needs in software~\cite{obaidi2025mood}. This allows project managers and requirements engineers to proactively address issues and foster more effective collaboration~\cite{patel2024sentimentgithub}.

To better interpret these moods or, for example, to identify negative moods at an early stage, sentiment analysis tools are employed~\cite{lin2018sentiment,zhang2025sentimentllm, sentianalyzerreport2022, sentisurvey-zenodo}. Sentiment analysis tools can be divided into lexicon-based approaches~\cite{islam2018sentistrengthse, herrmannSentiSurvey22} and those that utilize machine learning principles like large language models~\cite{ zhang2025sentimentllm}. A major challenge is that these tools are often not adapted to the SE domain, resulting in inaccurate outcomes~\cite{calefato2018senti, obaidi2021development, obaidiSentiSMS22}. Lin et al.~\cite{lin2018sentiment} cautioned researchers against relying on the results of sentiment analysis tools applied to SE-specific statements, as these tools are not yet mature enough for more advanced applications, such as recommending software libraries based on developer opinions.

While several gold-standard datasets for general sentiment analysis in German exist~\cite{cieliebak-etal-2017-twitter, momtazi2012fine}, none are specific to the software engineering (SE) domain. New sentiment analysis tools tailored to SE have been developed~\cite{calefato2018senti, islam2018sentistrengthse}, but, to the best of our knowledge, none have been trained on German data. As a result, a German gold-standard dataset is crucial for enabling the development of domain-specific and trustworthy~\cite{obaidi2025TrustworthySA} sentiment analysis tools for SE.

For these reasons, our work presents the creation of a gold-standard dataset using original German developer statements from the Android app development forum Android-Hilfe\footnote{\url{https://www.android-hilfe.de/}}.

In summary, this work makes the following contributions:
\begin{itemize}
    \item Construction of a gold-standard dataset with 5,949 German developer statements from the Android-Hilfe forum, annotated by multiple computer science students using a structured emotion model~\cite{shaver1987emotion}.
    \item Empirical evaluation of the dataset with four German sentiment analysis tools, highlighting current tool limitations in the SE domain.
\end{itemize}


\section{Background and Related Work}
\label{sec:background}

\subsection{Sentiment Analysis}

Sentiment analysis in software engineering (SE) has gained increasing research attention over the past decade~\cite{calefato2018senti, obaidi22cross, novielli2020cross, schroth2022potential, zhang2025sentimentllm}. Since software developers often work in teams and users post textual product reviews~\cite{obaidi2025automatingexplanationneedmanagement,obaidi2025AppFeaturesExplainNeeds,obaidi2025appKonwledge,anders2022userfeedback,anders2023userfeedback,obaidi2025elicit}, communication is frequent in this field~\cite{guzman2014appreviewsfeature}. Where human communication occurs, moods and emotions are present and can be detected~\cite{guzman2014appreviewsfeature}. Graziotin et al.~\cite{graziotin2014happy} showed that positive mood in developer teams improves cognitive performance, creativity, and analytical thinking. Therefore, analyzing moods and identifying trends can help maintain a positive team climate~\cite{graziotin2014happy}.


\subsection{Emotion Models}
Several emotion models describe emotions and are categorized into basic emotion, classification, and dimensional models~\cite{psychologie_ulich_mayring}. Dimensional models, such as Russell's two-dimensional framework~\cite{russell2003core} and the three-dimensional models by Mehrabian and Russell~\cite{mehrabian1974approach} and Osgood et al.~\cite{osgood1957measurement}, classify emotions along axes like valence, arousal, and dominance.

Basic emotion models define a set of fundamental emotions from which others can be derived. For instance, Izard~\cite{izard1992basic} identified ten basic emotions, and Plutchik~\cite{plutchik1982APT} proposed eight. These are considered to be present from birth and develop further over time~\cite{psychologie_ulich_mayring}.

Classification models, such as the hierarchical model by Shaver et al.~\cite{shaver1987emotion}, organize emotions into broad categories like \textit{Positive} and \textit{Negative}, with basic emotions (e.g., \textit{love, joy, anger, sadness, fear}) and their respective subcategories. While \textit{surprise} was initially considered as a basic emotion, Shaver et al. excluded it as a primary category due to its ambiguous valence and low empirical assignment.

Parrott~\cite{parrott2001emotions} extended Shaver et al.'s model by introducing a third, more detailed level, referring to six primary emotions, secondary emotions (as defined by Shaver et al.), and additional tertiary distinctions.


\subsection{Related Work}

\subsubsection{Creation of Gold-standard Datasets}

Several works have described their datasets as gold-standards, yet the methods for their creation vary. In the absence of universal guidelines, we summarize key approaches from related work in sentiment analysis to outline common practices.

Calefato et al.~\cite{calefato2018senti} developed the Senti4SD gold-standard dataset in a four-step process: (1) They reviewed emotion models and selected the concise Shaver et al.~\cite{shaver1987emotion} model to create a detailed annotation guideline. (2) Developer statements were extracted and cleaned of irrelevant content. (3) SentiStrength~\cite{islam2018sentistrengthse} was used to pre-balance the dataset across positive, negative, and neutral classes. Twelve computer science students annotated the data after a two-hour training, with statements labeled as “mixed” if containing emotions of different polarities. However, “mixed” labels were ultimately excluded from the gold-standard, and interrater agreement was ensured with a Cohen’s kappa of 0.74. (4) The final dataset was then used to develop the Senti4SD tool.

Similarly, Ortu et al.~\cite{ortu2016jira} constructed a gold-standard dataset from Jira, using Parrott’s extension of the Shaver model~\cite{parrott2001emotions}. Annotation was done by groups of students and researchers, with different group sizes and varying numbers of basic emotions used. Their dataset was imbalanced, with the majority of statements labeled as neutral~\cite{novielli2020cross}.

Novielli et al.~\cite{novielli2020cross} combined datasets from Stack Overflow, Jira, and 7,000 GitHub pull requests and commit comments\footnote{\url{https://doi.org/10.6084/m9.figshare.11604597.v1}}. All data were annotated with polarities \textit{Positive}, \textit{Negative}, and \textit{Neutral}, mapping the six basic emotions in the Jira dataset. These datasets contained only English developer statements.

Beyond SE, Saif et al.~\cite{saif2013verglichen_STS_gold_twitter} evaluated eight Twitter (now “X”) datasets for common issues such as inconsistent annotation, missing interrater agreement, and lack of entity-level analysis. Their own gold-standard dataset addressed these points by labeling each entity within a statement with \textit{Positive}, \textit{Negative}, \textit{Neutral}, \textit{Mixed}, or \textit{Other}, applying annotation guidelines and retaining only statements with high interrater reliability (yielding 2,200 of 3,000 statements), followed by further evaluation of the dataset’s accuracy.

Defining a gold-standard thus proves challenging, as researchers often apply their own interpretations. However, some commonalities can be identified. Novielli et al.~\cite{novielli2020cross} noted that subjective label assignment without a guideline leads to ambiguous gold-standard datasets. It is therefore important to define a clear annotation guideline based on a theoretical emotion model and to provide sufficient training for raters~\cite{novielli2020cross, saif2013verglichen_STS_gold_twitter}. Furthermore, datasets should not be labeled by a single person to avoid subjectivity, although Ortu et al.~\cite{ortu2016jira} found that agreement remained constant with more than two raters. Novielli et al.~\cite{novielli2020cross} also emphasized that high interrater reliability is crucial for the validity of a gold-standard dataset, making its calculation essential.

\subsubsection{German Datasets}

Outside SE, Boland et al.~\cite{boland2013creating} created a German dataset of Amazon\footnote{\url{https://amazon.com/}} reviews, where over 63,000 sentences were annotated with polarities by nine raters.

Sänger et al.~\cite{sanger-etal-2016-scare} created a German data corpus called SCARE\footnote{\url{https://www.romanklinger.de/scare/}} using app reviews from the Google Play Store, collecting 1,760 different reviews, which they further divided into 3,953 subjective sentences. The reviews were drawn from apps in eleven different categories. The resulting dataset was rather imbalanced, with about 62\% of the reviews labeled as \textit{Positive}, 36\% as \textit{Negative}, and only 1.6\% as \textit{Neutral}~\cite{sanger-etal-2016-scare}. The annotation process was not based on an emotion model.



\section{Creation of Dataset}
\label{sec:research}

\subsection{Crawling}

\subsubsection{Selection of Source}
Since developer forums such as Stack Overflow or GitHub contain predominantly English-language statements and were thus unsuitable for this study, a German developer forum was sought. Based on criteria such as language, relevance to the software engineering domain, feasibility of data extraction, and the volume of available data, the Android app development section of the German forum Android-Hilfe\footnote{\url{https://www.android-hilfe.de/forum/android-app-entwicklung.9/}} was selected. As of June 15, 2022, this forum contains 14,088 topics and 74,946 posts, offering a rich source of German content. Furthermore, the forum rules\footnote{\url{https://www.android-hilfe.de/help/regeln/}} require that all contributions be written exclusively in German and that users should strive for correct spelling. With an average of approximately 3.4 posts per topic, the forum demonstrates a high level of communication among developers. 

\subsubsection{Crawler Functionality}
The crawler for the Android-Hilfe.de developer forum was implemented using the Python framework Scrapy\footnote{\url{https://scrapy.org/}}, an open-source tool for extracting web data. The crawler starts from the forum’s first page and iterates through each thread, sending every thread to the \texttt{crawl\_posts} function, where all posts on that page are processed. Due to the forum’s page structure, individual lines from posts must be concatenated. Images, which are irrelevant for this work and indicated by the text ``Click to enlarge...'', are filtered out. To ensure that posts are not too long, a limit of 200 characters was set. Additionally, statements containing quotes or automatically generated text were excluded. Once all posts in a thread have been processed, the crawler uses the ``Next Page'' button to continue to the following page. The entire process is recursive and continues until the last page has been visited.

\subsubsection{Dataset Composition}
The dataset, generated with the Android-Hilfe crawler, comprises 20,380 unique German developer statements. To ensure a balanced dataset, we performed preliminary sorting using the sentiment analysis tool GerVADER~\cite{tymann2019gervader}, which identified 10,560 positive, 7,046 neutral, and 2,774 negative statements. Since GerVADER assigns sentiment scores within the interval of -1 (negative) to +1 (positive), the statements were sorted in descending or ascending order accordingly. This is important to minimize contradictory statements that could later be assigned to different emotions or polarities. Subsequently, the 2,000 statements with the highest scores for each of the three polarities were selected and merged to form a dataset of 6,000 statements. Some of these included irrelevant information such as ``sent from my iPhone XR'' (translated to English), which users could enable as a signature. Such signatures were manually removed in a post-processing step.

\subsection{Labeling}
\label{sec:labeling}
\subsubsection{Guideline for the Labeling Process}

For the labeling process, we followed related work~\cite{novielli2018stackoverflow-gold,novielli2020cross} and used the basic emotions proposed by Shaver et al.\cite{shaver1987emotion}. The emotions \textit{Love}, \textit{Joy}, \textit{Surprise}, \textit{Anger}, \textit{Sadness}, and \textit{Fear} were included.

In addition to the simplified emotion model by Shaver et al.~\cite{shaver1987emotion}, the basic emotion \textit{Surprise} was incorporated, as it is also considered in other gold-standard datasets~\cite{novielli2018stackoverflow-gold}, and Shaver et al. themselves continued to treat \textit{Surprise} as a basic emotion outside the main diagram~\cite{shaver1987emotion}. Similar to the guidelines by Novielli et al.~\cite{novielli2018stackoverflow-gold}, raters were instructed to note all emotions identified in a statement, but to agree on a single emotion to ensure a single label for each statement. If no emotion could be identified, the statement was to be labeled as \textit{Neutral}.

\subsubsection{Workshop Participants}

Five computer science students participated in the labeling workshop. All participants were male and between 20 and 25 years old. One participant had already completed their bachelor's degree, while three others, were writing their theses at the time of the workshop. All five participants had prior experience in software development and had worked in developer teams with regular communication.

\subsubsection{Workshop Procedure}

The five participants met to jointly review the Shaver et al.\cite{shaver1987emotion} guideline. To prevent possible ambiguities that might arise later, a sample dataset containing 20 statements (not included in the original dataset) was introduced. Once all uncertainties were clarified, each participant received their subset of 3,000 statements to label, while one participant received the complete set of 6,000 statements. This resulted in two groups.

As an initial step, the participants labeled 100 statements each. Afterward, the raters within each group met to compare their labels. This step was intended to uncover discrepancies before labeling the larger remainder of the dataset and to identify areas of disagreement and their causes. Additionally, participants were able to clarify any questions.

The first inconsistencies were discussed and are described in Section~\ref{erster_durchgang}. After the workshop, each participant was given four weeks to complete their assigned labeling, during which they also revised the first 100 labels based on the issues discussed in Section~\ref{erster_durchgang}. One participant from the second group discontinued labeling due to time constraints, so a member from the first group took over their full subset. Once all labeling was complete, the groups met again to discuss inconsistencies and their resolutions. These are described in Section~\ref{letzter_durchgang}.

\subsection{Data Analysis}
\label{sec:data_analysis}
\subsubsection{Performance Metrics}
\label{sec:metrics}
To evaluate the performance of different sentiment analysis tools, we computed precision, recall, the micro- and macro-averaged $F_1$ scores, as well as their mean (overall score), following established practices~\cite{novielli2020cross,9240704}. Each statement was assigned a single polarity label, making the micro $F_1$ score equivalent to accuracy. The $F_1$ score balances precision and recall, offering a robust metric for overall performance. Macro $F_1$ averages scores across classes equally, while micro $F_1$ weights them by class size. Both provide complementary insights into tool performance.

\subsubsection{Interrater Agreement}
\label{sub:interrater}
To calculate the agreement between the three raters (three researchers), we computed Fleiss' $\kappa$. Fleiss' $\kappa$ quantifies the interrater reliability in scenarios involving more than two raters. For pairwise comparisons, we also report Cohen's $\kappa$, which measures agreement between two raters. Both $\kappa$ values are interpreted according to the scheme outlined by Landis and Koch~\cite{landis1977measurement}. In addition, we provide the agreement in percentage. For example, if three raters agree in six out of 10 cases, then the agreement is 60\%.


\section{Results}
\label{sec:results}
The crawler software and the dataset including all calculations and analysis results are openly available at \href{https://doi.org/10.5281/zenodo.15844091}{Zenodo}~\cite{obaidi2025datasetsentimentgold}.

\subsection{Evaluation of the Labeling Process}

To evaluate the labeling process, we first examine the disagreements between the raters during label assignment. The results of this process are then presented.

\subsubsection{Disagreements after the First Round}
\label{erster_durchgang}

After the first round of labeling, in which the initial 100 statements were evaluated, it became evident that there were difficulties in assigning the emotion \textit{Love} correctly. This issue was resolved by referring to the English gold-standard dataset by Novielli et al.~\cite{novielli2018stackoverflow-gold}, where \textit{Love} is assigned when a person expresses a positive attitude toward another person, for example, by offering praise or thanks. In the latter case, there was also some uncertainty, as an expression of thanks was interpreted by some participants as \textit{Joy}. These cases served as prime examples of the need to judge statements as objectively as possible and to assign a label only when the speaker explicitly expresses an emotion through their wording. This principle also helped to resolve disagreements regarding the assignment of other emotions, as some participants had initially allowed too much subjectivity to influence their labeling.

This tendency is illustrated by the following example, translated from German to English:
\begin{quote}
    ``That doesn't work either. There is always an error message: so how am I supposed to do it now?''
\end{quote}
Two raters detected the emotion \textit{Sadness}, while the third assigned \textit{Fear}. The raters explained their choices by saying that in such a situation, they would personally feel desperate or anxious. However, since the statement itself does not explicitly express any emotion, it must be classified as neutral. In general, participants agreed that insults should be categorized as \textit{Anger}. Furthermore, questions need not be assigned the emotion \textit{Surprise}; in such cases, signal words like ``strange'' or ``weird'' better indicate an emotion.

It was observed that raters sometimes forgot to assign \textit{Surprise} with an additional \textit{Positive} or \textit{Negative} polarity, mainly due to its rarity and oversight. Therefore, participants were reminded to correct this in their datasets and be mindful in future annotations.

\subsubsection{Disagreements after the Final Round}
\label{letzter_durchgang}

After all participants had completed labeling, the two groups met separately to resolve disagreements and produce a final dataset with unambiguous emotion labels. An initial analysis revealed that, in 1,205 out of 6,000 cases, there was at least one rater who assigned a different label than the other two. Among these 1,205 cases, only 91 involved all three raters assigning different labels. These disagreements were discussed in group meetings to determine a final label for each case.

An example of such a disagreement can be observed in the following developer statement, translated from German to English:
\begin{quote}
    ``Too bad, but I had already feared something like this. Well, then I just have to struggle with endlessly long file names.''
\end{quote}
Rater 1 assigned the emotion \textit{Fear}, as the word ``feared'' was seen as a signal for this emotion. Rater 2 identified \textit{Sadness} due to the phrase ``too bad'', which clearly expresses a feeling. Rater 3 saw \textit{Anger}, as the use of ``struggle'' was interpreted as expressing annoyance and thus an aggressive tone.

After a five-minute discussion, the raters agreed to assign the label \textit{Sadness}, since ``feared'' was considered to refer to a past emotion and was not relevant to the current emotional state expressed in the statement. Furthermore, ``struggle'' does not necessarily indicate \textit{Anger}. Ultimately, the key word ``too bad'' prevailed, justifying the assignment of \textit{Sadness}.

In some cases, raters were surprised by the labels they had assigned and quickly agreed with another rater, often explaining that they had pressed the wrong key by mistake and had simply overlooked it. It was also agreed that statements (51 in total) in English or those consisting solely of code would be removed from the dataset. The process resulted in a dataset of 5,949 manually labeled developer statements. 

\subsection{Results of the Final Dataset}
\label{sec:results_data}

The following presents the initial evaluation of the final dataset. Figure~\ref{fig:zsm} shows the composition of dataset polarities across the three different phases.

\begin{figure}[!htb]
 \centering
 \includegraphics[width=1\linewidth]{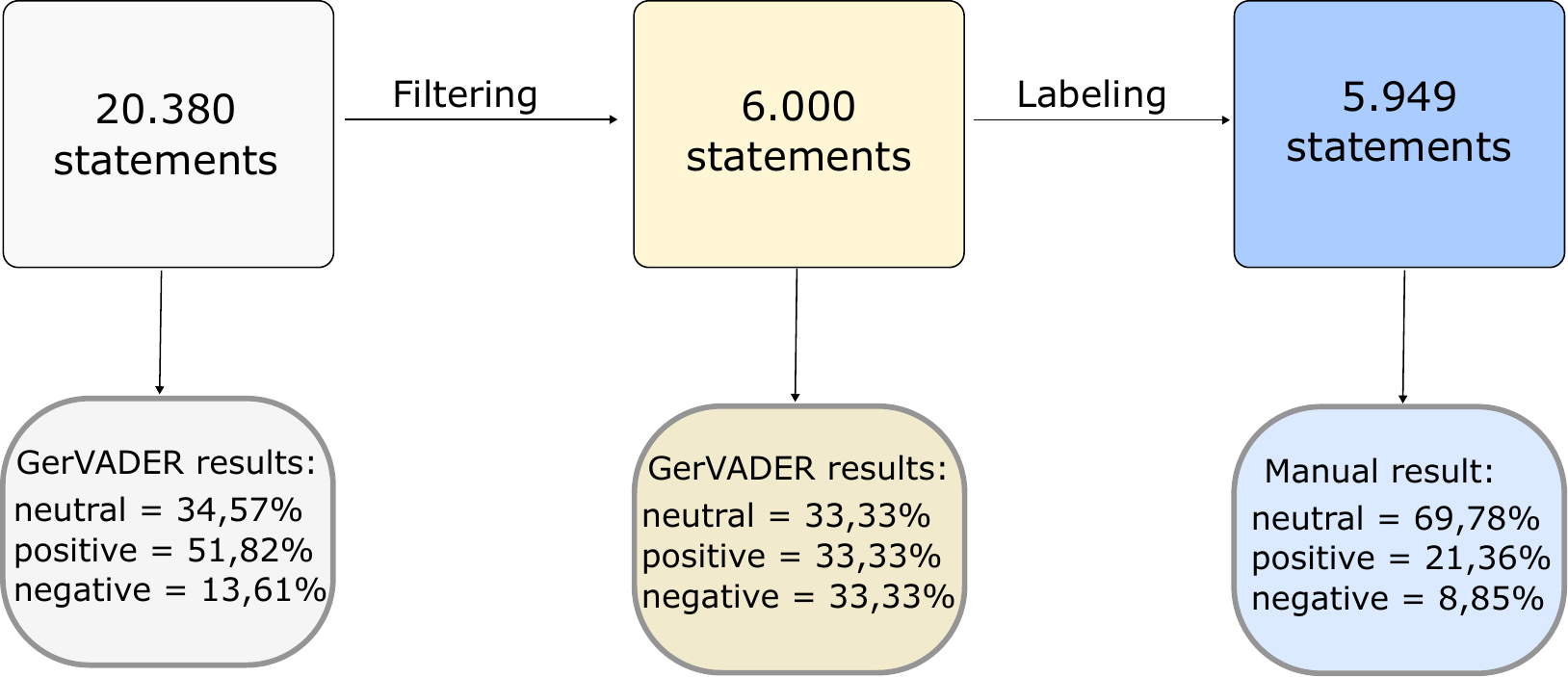}
 \caption{Composition of the dataset at different phases}
 \label{fig:zsm}
\end{figure}

The first evaluation of the labeling process showed that the dataset of 5,949 statements consists of 69.78\% neutral, 21.36\% positive, and 8.85\% negative developer statements. These values were calculated from the entries in Table~\ref{tab:vorkommen-datensatz}, following the same translation from emotions to polarities as described by Novielli et al.~\cite{novielli2020cross}: \textit{Love}, \textit{Joy}, and \textit{Positive Surprise} were mapped to the polarity \textit{Positive}, while \textit{Negative Surprise}, \textit{Anger}, \textit{Sadness}, and \textit{Fear} were mapped to \textit{Negative}.

\begin{table*}[htb]
\centering
{{\begin{tabular}{|c|cccccccc|c}
\hline
\multirow{2}{*}{\textbf{Statements}} & \multicolumn{8}{c|}{\textbf{Statements with this label}}                                                                                                                                     & \multicolumn{1}{c|}{\multirow{2}{*}{\textbf{N}}} \\ \cline{2-9}
                                       & \multicolumn{1}{c|}{Neutral} & \multicolumn{1}{c|}{Love}   & \multicolumn{1}{c|}{Joy} & \multicolumn{1}{c|}{Pos. Surprise} & \multicolumn{1}{c|}{Neg. Surprise} & \multicolumn{1}{c|}{Anger} & \multicolumn{1}{c|}{Sadness} & Fear  & \multicolumn{1}{c|}{}                    \\ \hline
\#                                     & \multicolumn{1}{c|}{4,151}   & \multicolumn{1}{c|}{1,134}  & \multicolumn{1}{c|}{133} & \multicolumn{1}{c|}{4}            & \multicolumn{1}{c|}{46}           & \multicolumn{1}{c|}{89}    & \multicolumn{1}{c|}{384}     & 8     & \multicolumn{1}{c|}{5,949}               \\ \hline
\%                                     & \multicolumn{1}{c|}{69.78\%} & \multicolumn{1}{c|}{19.06\%}& \multicolumn{1}{c|}{2.24\%}& \multicolumn{1}{c|}{0.07\%}         & \multicolumn{1}{c|}{0.77\%}        & \multicolumn{1}{c|}{1.5\%} & \multicolumn{1}{c|}{6.45\%}  & 0.13\%&                                    \\ \cline{1-9}
\end{tabular}}
\caption{\label{tab:vorkommen-datensatz}Distribution of emotion labels in the dataset}}
\end{table*}

The most frequent emotions were \textit{Love} (19.06\%) and \textit{Sadness} (6.45\%), while \textit{Positive Surprise} (0.07\%) and \textit{Fear} (0.13\%) were least frequent.

\begin{table*}[htb]
\centering
\begin{tabular}{|c|c|c|c|c|cc|c|c|c|c|}
\hline
\textbf{Round} &
  \textbf{Metric} &
  \textbf{Neutral} &
  \textbf{Love} &
  \textbf{Joy} &
  \multicolumn{1}{c|}{\textbf{Pos. Sur.}} &
  \textbf{Neg. Sur.} &
  \textbf{Anger} &
  \textbf{Sadness} &
  \textbf{Fear} &
  \textbf{Total} \\ \hline
\multirow{2}{*}{1} & Agreement  & 0.63 & 0.86 & 0.71 & \multicolumn{2}{c|}{0.88}        & 0.84 & 0.9  & 0.92  & 0.43 \\ 
                    & Fleiss' K & 0.5  & 0.04 & 0.32 & \multicolumn{2}{c|}{0.23}        & 0.33 & 0.39 & 0.03  & 0.36 \\ \hline
\multirow{2}{*}{2} & Agreement  & 0.82 & 0.93 & 0.96 & \multicolumn{1}{c|}{0.98} & 0.99 & 0.96 & 0.94 & 0.995 & 0.8  \\
                    & Fleiss' K & 0.71 & 0.85 & 0.47 & \multicolumn{1}{c|}{0.28} & 0.1  & 0.37 & 0.67 & 0.34  & 0.71 \\ \hline
\multirow{2}{*}{Diff.} &
  Agreement &
  \multicolumn{1}{l|}{+.19} &
  \multicolumn{1}{l|}{+.07} &
  \multicolumn{1}{l|}{+.25} &
  \multicolumn{2}{c|}{\multirow{2}{*}{-}} &
  \multicolumn{1}{l|}{+.12} &
  \multicolumn{1}{l|}{+.04} &
  \multicolumn{1}{l|}{+.08} &
  \multicolumn{1}{l|}{+.37} \\
 &
  Fleiss' K &
  \multicolumn{1}{l|}{+.21} &
  \multicolumn{1}{l|}{+.81} &
  \multicolumn{1}{l|}{+.15} &
  \multicolumn{2}{c|}{} &
  \multicolumn{1}{l|}{+.04} &
  \multicolumn{1}{l|}{+.28} &
  \multicolumn{1}{l|}{+.31} &
  \multicolumn{1}{l|}{+.35} \\ \hline
\end{tabular}
\caption{\label{tab:Reliabilitäten}Interrater agreement (Fleiss' Kappa) for emotion labels. In the first annotation round, raters did not distinguish between positive and negative surprise; thus, only a combined value for surprise is reported.}

\end{table*}

To assess the labeling validity and the impact of intermediate discussions on agreement and interrater reliability, these metrics were calculated as shown in Table~\ref{tab:Reliabilitäten}.

Moreover, it can be seen that, although the agreement rates are quite high across emotions, the Fleiss' Kappa values are relatively low. Both agreement and Fleiss' Kappa for all labels increased after the second labeling round. Particularly high agreement rates were achieved for the emotions \textit{Fear}, \textit{Negative Surprise}, and \textit{Positive Surprise}, although these also had the lowest Fleiss' Kappa values. The Fleiss' Kappa for \textit{Love} showed a particularly large increase, with a difference of +0.81. Both the overall agreement and Fleiss' Kappa for all emotions increased by +0.37 and +0.35, respectively. Novielli et al.~\cite{novielli2018stackoverflow-gold} reported very similar agreement and Fleiss' Kappa values for individual emotions. However, their Kappa values for \textit{Fear} and \textit{Love} were considerably lower at 0.45 and 0.66, compared to 0.67 and 0.85 in our results, while \textit{Anger} had a much lower value in our study (0.37 vs. 0.62). The other emotions showed only minor deviations~\cite{novielli2018stackoverflow-gold}.

Table~\ref{tab:Reliabilitäten_pol} shows the interrater agreement for the polarities translated from the emotions. Here, too, all values increased in the second round, although the increase was more moderate since the initial values were already high.

\begin{table}[htb]
\centering
\begin{tabular}{|c|c|c|c|c|c|}
\hline
\textbf{Round} & \textbf{Metric} & \textbf{Neutral} & \textbf{Positive} & \textbf{Negative} & \textbf{Overall} \\ \hline
\multirow{2}{*}{1}    & Agreement  & 0.67 & 0.83 & 0.78 & 0.66 \\
                       & Fleiss' K & 0.55 & 0.69 & 0.51 & 0.58 \\ \hline
\multirow{2}{*}{2}    & Agreement  & 0.82 & 0.91 & 0.89 & 0.81 \\
                       & Fleiss' K & 0.71 & 0.81 & 0.59 & 0.73 \\ \hline
\multirow{2}{*}{Diff.} & Agreement  & +.15 & +.08 & +.11 & +.15 \\
                       & Fleiss' K & +.16 & +.12 & +.08 & +.15 \\ \hline
\end{tabular}
\caption{\label{tab:Reliabilitäten_pol}Interrater agreement for polarities}
\end{table}

\subsection{Evaluation Using Sentiment Analysis Tools}

For the evaluation with sentiment analysis tools, four German tools were selected, and their results are presented below.

\subsubsection{Selection of Tools}

For evaluating the dataset with German sentiment analysis tools, the lexicon-based GerVADER~\cite{tymann2019gervader}, SentiStrength\_DE\footnote{\url{https://github.com/OFAI/SentiStrength_DE}}, and TextBlobDE\footnote{\url{https://github.com/markuskiller/textblob-de}}, as well as the machine-learning tool BertDE~\cite{guhr2020training}, were chosen.

\subsubsection{Tool Results}

To evaluate the created dataset with these sentiment analysis tools, precision, recall, and $F_1$-scores for the three polarity classes were calculated, as well as macro averages and an accuracy value.

The results in Table~\ref{tab:toolsauswertung} clearly show that the sentiment analysis tool SentiStrength\_DE achieved the highest macro-average $F_1$-scores and accuracy values.
\begin{table*}[htb]
\centering
\begin{tabular}{|c|ccc|ccc|ccc|ccc|c|}
\hline
\multirow{2}{*}{\textbf{Tool}} &
  \multicolumn{3}{c|}{\textbf{Neutral}} &
  \multicolumn{3}{c|}{\textbf{Positive}} &
  \multicolumn{3}{c|}{\textbf{Negative}} &
  \multicolumn{3}{c|}{\textbf{Macro-avg}} &
  \multirow{2}{*}{\textbf{Accuracy}} \\ \cline{2-13}
                    & P   & R   & $F_1$  & P   & R   & $F_1$  & P   & R   & $F_1$  & P   & R   & $F_1$  &     \\ \hline
\textbf{GerVADER}   & \textbf{.89} & .57 & .70 & .53 & \textbf{.84} & \textbf{.65} & .24 & .58 & .34 & .55 & \textbf{.67} & .56 & .63 \\ \hline
\textbf{BertDE}     & .88 & .22 & .35 & .60 & .61 & .61 & .13 & \textbf{.90} & .23 & .54 & .58 & .39 & .36 \\ \hline
\textbf{\begin{tabular}[c]{@{}c@{}}Senti-\\ Strength\_DE\end{tabular}} &
  .78 &
  \textbf{.85} &
  \textbf{.82} &
  \textbf{.67 }&
  .39 &
  .49 &
  \textbf{.38} &
  .47 &
  \textbf{.42} &
  \textbf{.61} &
  .57 &
  \textbf{.58} &
  \textbf{.72} \\ \hline
\textbf{TextBlobDE} & .72 & .70 & .71 & .35 & .37 & .36 & .15 & .17 & .16 & .41 & .41 & .41 & .58 \\ \hline
\end{tabular}
\caption{Performance of the dataset in relation to the tools. Highest values are bold.}
\label{tab:toolsauswertung}
\end{table*}
Thus, SentiStrength\_DE achieves the highest accuracy on the created dataset with an accuracy of 0.72, though this still does not represent strong performance. BertDE achieves the lowest accuracy with 0.36.

Compared to results reported in the systematic mapping study by Obaidi et al.~\cite{obaidiSentiSMS22}, the performance of English lexicon-based tools such as SentiStrength~\cite{thelwall2010sentiment} and DEVA~\cite{islam2018deva} (accuracy: 0.71 and 0.77; macro-average $F_1$: 0.61 and 0.75) is noticeably higher than that of the German lexicon-based tools on our dataset. Only SentiStrength\_DE, with an accuracy of 0.72 and a macro-average $F_1$-score of 0.58, achieves comparably high results.

For comparison with the machine learning tool BertDE, one can consider the performance values of English machine-learning-based tools such as Senti4SD~\cite{calefato2018senti} or SentiCR~\cite{ahmed2017senticr} listed in the systematic mapping study by Obaidi et al.~\cite{obaidiSentiSMS22}. These were also trained on data from other domains. On average, they achieved accuracy values of 0.74 and 0.77 and macro-average $F_1$-scores of 0.66 and 0.69, respectively. In comparison, BertDE scored much lower on the German gold-standard dataset created here, with an accuracy of 0.36 and a macro-average $F_1$-score of 0.39.

SentiStrength\_DE achieves also the highest $F_1$-score for the \textit{Neutral} class with a value of 0.82. It is noteworthy that the other lexicon-based tools also show elevated values for this class. With an $F_1$-score of 0.16 for \textit{Negative}, TextBlobDE shows the lowest value overall. In general, $F_1$-scores for the \textit{Negative} class are quite low across all tools. To some extent, these results are reflected in the interrater agreements between the dataset and each sentiment analysis tool(cf. Table~\ref{tab:cohenskappa}). 

GerVADER and SentiStrength\_DE achieve the highest Cohen's Kappa values, with 0.38 and 0.35 respectively, but these values are still to be considered minimal.

\begin{table}[htb]
\centering
\begin{tabular}{|c|c|c|c|c|c|}
\hline
\textbf{Tool}                                                          & \textbf{Metric} & \textbf{Neutral} & \textbf{Positive} & \textbf{Negative} & \textbf{Overall} \\ \hline
\textbf{GerVADER}   & Cohen’s K & \textbf{0.37} & \textbf{0.52} & 0.25 & \textbf{0.38} \\ \hline
\textbf{BertDE}     & Cohen’s K & 0.10          & 0.50          & 0.09 & 0.18          \\ \hline
\textbf{\begin{tabular}[c]{@{}c@{}}Senti-\\ Strength\_DE\end{tabular}} & Cohen’s K  & 0.32             & 0.40             & \textbf{0.36}    & 0.35            \\ \hline
\textbf{TextBlobDE} & Cohen’s K & 0.08          & 0.18          & 0.07 & 0.12          \\ \hline
\end{tabular}
\caption{\label{tab:cohenskappa} Interrater agreement between human annotation and sentiment analysis tools. Highest values are bold.}
\end{table}

\section{Discussion}
\label{sec:discussion}

In the following, we interpret the results, present threats to validity and future work.

\subsection{Interpretation}
\label{sec:interpretation}

The results from Section~\ref{sec:results} show that, despite some disagreements between raters after both rounds of labeling, high agreement and reliability values were achieved. These values closely resemble those found in other gold-standard datasets in the SE domain, such as those by Novielli et al.~\cite{novielli2020cross, novielli2018stackoverflow-gold}. As high interrater reliability is recognized as a key criterion for dataset validity~\cite{novielli2020cross}, this work succeeded in achieving the goal of create a German gold-standard dataset for sentiment analysis of developer statements.

The differences in these values between the two labeling rounds highlight that an intermediate discussion to resolve disagreements is indispensable for producing a robust gold-standard dataset. Especially the substantial increase in the Kappa value for the emotion \textit{Love} in Table~\ref{tab:Reliabilitäten} demonstrates that, even with clear guidelines, every person introduces a certain level of subjectivity when assigning emotion labels. An intermediate review can minimize such differences. When considering the polarities in Table~\ref{tab:Reliabilitäten_pol}, the improvements due to the intermediate discussion are not as pronounced, since distinguishing between \textit{Positive} and \textit{Negative} is more intuitive than deciding on a specific basic emotion. Nevertheless, the discussion is helpful here, especially for assigning statements to the \textit{Neutral} class, ensuring potential rater subjectivity is separated from the actual sentiment expressed.

Despite pre-sorting the dataset with GerVADER~\cite{tymann2019gervader}, the resulting dataset is still imbalanced, with 69.78\% neutral statements (see Table~\ref{tab:vorkommen-datensatz}). The low $F_1$-scores for polarities, when compared to human-assigned labels in Table~\ref{tab:toolsauswertung}, indicate that GerVADER tends to detect polarity even in neutral developer statements, particularly for the \textit{Negative} class. Since SentiStrength\_DE achieved the highest accuracy at 0.72, it may be better suited than GerVADER for pre-sorting German sentences in the SE domain. However, this may vary depending on the data source, as developer statements are inherently subjective expressions by different individuals.

Although SentiStrength\_DE achieves similarly high macro-average $F_1$-scores and accuracy as comparable English tools such as SentiStrength, when considering averages over multiple datasets~\cite{obaidiSentiSMS22}, and would therefore appear to be the best suited for the domain used in this work, both its scores and Cohen’s Kappa values remain unsatisfactory (see Table~\ref{tab:cohenskappa}). It should be noted that Imtiaz et al.~\cite{imtiaz18sentiment} found that no tool is fully reliable for the SE domain, and that human and tool rarely reach strong agreement. Thus, these tools should not be solely relied upon, but may serve as a starting point for future evaluations in this area.
This shows the need for a high-quality, domain-specific gold-standard dataset and dedicated reliable sentiment analysis tools for the SE context, especially in German, so that sentiment trends in developer communication can be assessed more accurately and leveraged for tasks such as early detection of issues, prioritization of requirements, and effective management in CrowdRE scenarios.

When examining the accuracy and $F_1$-scores of the machine-learning-based tool BertDE, it is apparent that this tool performs worst. It should be noted that BertDE was not trained for the SE domain. The English tool Senti4SD~\cite{calefato2018senti}, for example, also performs poorly when evaluated on out-of-domain datasets~\cite{zhang20sentiment}, but achieves much better results (accuracy of 0.89) when applied to datasets from the same domain. Ben et al.~\cite{ben2010theory} noted that domain specificity is crucial for machine learning to yield accurate results. Thus, it is possible that BertDE would also perform well if trained on the dataset created in this study.

To further confirm the validity of the dataset in the SE domain, it would have been beneficial if existing German sentiment analysis tools had performed well in this domain. Since this was not the case, it is important to develop established German sentiment analysis tools specifically for SE, as English-language tools in this area achieve better performance~\cite{obaidiSentiSMS22}. As no such German tool currently exists, the dataset created here could be used for its development.

\subsection{Future Work}
\label{sec:ausblick}

The concept of dataset creation and its results presented in this work may assist in the development of further datasets. Future work should investigate to what extent generative large language models (LLMs) can be used for filtering datasets, and how their performance compares to more traditional methods in sentiment analysis. 
To minimize the risk of raters making input errors during labeling, the use of a new concept or dedicated software may be beneficial. For evaluation with sentiment analysis tools, it is important to ensure they are tailored to the specific domain. Moreover, using more sentiment analysis tools during evaluation helps produce more representative results. 
In constructing a German machine-learning-based sentiment analysis tool, the dataset developed in this work could be used for training. However, such tools generally achieve higher accuracy as the size of the gold-standard dataset increases. Therefore, it would be necessary to create additional datasets from other German sources, while considering the optimizations proposed in this study.

\subsection{Threats to Validity}
\label{sec:validiteat}

We structure threats to validity following Wohlin et al.~\cite{wohlin2012experimentation}:

\textbf{Construct validity.}  
A key limitation is that, for the evaluation and comparison of our results, we had to rely on findings from English-language studies, as no comparable gold-standard dataset for German developer statements in the SE domain currently exists. Consequently, our comparisons are subject to potential linguistic and cultural differences between German and English communication. Furthermore, the number of sentiment analysis tools evaluated is limited, although we used the most widely established, freely available tools for German sentiment analysis. Another potential threat arises from the data source: although Android-Hilfe.de seems suitable for collecting authentic developer communication, Lin et al.~\cite{lin2018sentiment} showed that training on inadequate sources can harm results, as seen in prior work with Stack Overflow.

\textbf{Internal validity.}  
Human annotation is inherently imperfect: raters can make mistakes such as typos or lose concentration during lengthy labeling sessions. Although all statements in the final dataset were labeled by three different raters, and majority voting was used to select the final label, it remains possible that at least two raters made an incorrect decision for a given statement, leading to errors in the final dataset. Additionally, contrary to our original plan, one rater from the first group labeled an extra 3,000 statements due to another rater’s withdrawal. In the evaluation of the first round, the labels from the withdrawn rater were still included, potentially introducing bias in round-to-round comparisons. However, all statements were rated by three different raters in the final dataset, and the positive effects of intermediate discussion on agreement still hold.

\textbf{Conclusion validity.}  
The annotation process involved only three raters per statement, which might be considered a limitation since larger groups of raters are often recommended for more robust results. Nevertheless, prior work by Ortu et al.~\cite{ortu2016jira} found that increasing the number of raters beyond two does not significantly affect label assignment. We also observed considerable fluctuations in agreement and Fleiss' Kappa values, indicating that full stability among raters was not always achieved and that further intermediate discussions could have led to additional changes. These fluctuations were mostly observed within the polarity classes.

\textbf{External validity.}  
All raters in this study were male and between 20 and 25 years old. It remains an open question whether a more diverse group of raters, considering gender, age, or background, would yield different results. As such, the generalizability of the dataset and our findings to other groups should be interpreted with caution.

\section{Conclusion}
\label{sec:conclusion}

Sentiment analysis in the software engineering (SE) domain is used to support positive moods and to identify negative sentiments early, as studies have shown that positive sentiment among developers increases productivity. While sentiment analysis tools tailored to SE exist, they are currently only available for English-language content. Therefore, this work aimed to create a German gold-standard dataset of developer statements to enable the development of a machine learning-based sentiment analysis tool for the SE domain in German. To this end, we selected a high-quality data source and developed a clear annotation guideline, based on established research. The evaluation showed that, in addition to a strong guideline, intermediate discussions among raters are essential for achieving high agreement and reliability. The resulting dataset consists of 5,949 German developer statements, each labeled with one of six basic emotions according to Shaver et al.~\cite{shaver1987emotion} by three raters. The high agreement and reliability values indicate that this dataset can be considered a representative gold-standard. Evaluations with existing German sentiment analysis tools showed that their performance is inadequate for the SE domain, confirming the need for a dedicated German sentiment analysis tool. The gold-standard dataset created in this study provides a valuable foundation for this purpose.


\bibliographystyle{IEEEtran}
\bibliography{references.bib}

\end{document}